\documentclass[twocolumn,showpacs,preprintnumbers,amsmath,amssymb,superscriptaddress,aps,prb, standalone,floatfix,longbibliography]{revtex4-1}
\usepackage{amsmath}
\usepackage{xcolor}
\usepackage{braket}
\usepackage{gensymb}
\usepackage{graphicx}
\usepackage{dcolumn}
\usepackage{bm}

\usepackage[utf8]{inputenc}
\usepackage[T1]{fontenc}
\usepackage{mathptmx}
\usepackage{etoolbox}

\begin{document}

\title{Demonstration of NV-detected $^{13}$C NMR at 4.2 T}

\author{Yuhang Ren}
\affiliation{Department of Physics and Astronomy, University of Southern California, Los Angeles, CA 90089, USA}

\author{Cooper Selco}
\affiliation{Department of Physics and Astronomy, University of Southern California, Los Angeles, CA 90089, USA}

\author{Dylan Kawashiri}
\affiliation{Department of Physics and Astronomy, University of Southern California, Los Angeles, CA 90089, USA}

\author{Michael Coumans}
\affiliation{Department of Chemistry, University of Southern California, Los Angeles, CA 90089, USA}

\author{Benjamin Fortman}
\affiliation{Department of Chemistry, University of Southern California, Los Angeles, CA 90089, USA}

\author{Louis-S. Bouchard}
\affiliation{Department of Chemistry \& Biochemistry, University of California, Los Angeles CA 90095, USA}

\author{Karoly Holczer}
\affiliation{Department of Physics \& Astronomy, University of California, Los Angeles CA 90095, USA}

\author{Susumu Takahashi}
\email{susumu.takahashi@usc.edu}
\affiliation{Department of Chemistry, University of Southern California, Los Angeles, CA 90089, USA}
\affiliation{Department of Physics and Astronomy, University of Southern California, Los Angeles, CA 90089, USA}

\date{\today}

\begin{abstract}
The nitrogen-vacancy (NV) center in diamond has enabled studies of nanoscale nuclear magnetic resonance (NMR) and electron paramagnetic resonance with high sensitivity in small sample volumes. Most NV-detected NMR (NV-NMR) experiments are performed at low magnetic fields. While low fields are useful in many applications, high-field NV-NMR with fine spectral resolution, high signal sensitivity, and the capability to observe a wider range of nuclei is advantageous for surface detection, microfluidic, and condensed matter studies aimed at probing micro- and nanoscale features. However, only a handful of experiments above 1 T were reported. Herein, we report $^{13}$C NV-NMR spectroscopy at 4.2 T, where the NV Larmor frequency is 115 GHz. Using an electron-nuclear double resonance technique, we successfully detect NV-NMR of two diamond samples. The analysis of the NMR linewidth based on the dipolar broadening theory of Van Vleck shows that the observed linewidths from sample 1 are consistent with the intrinsic NMR linewidth of the sample. For sample 2 we find a narrower linewidth of 44 ppm. This work paves the way for new applications of nanoscale NV-NMR.
\end{abstract}

\maketitle

\section{Introduction}
A nitrogen-vacancy (NV) center is a fluorescent impurity center found in the diamond lattice~\cite{Loubser1978}.
The NV center is one of the most promising tools for fundamental research in quantum information science and its applications because of its unique properties, including the capability of individual NV detection through optically detected magnetic resonance (ODMR) spectroscopy, a high degree of spin polarizability with optical excitation, and long coherence times from room temperature down to cryogenic temperatures~\cite{Gruber1997, Jelezko2004, Childress2006, Epstein2005, Gaebel2006, Takahashi2008, Balasubramanian2009}.
The NV center is an excellent quantum sensor~\cite{Degen2008, Balasubramanian2008, Maze2008, Taylor2008}.
It not only has demonstrated unprecedented magnetic field sensitivity over a wide temperature range but also shows nanometer-scale spatial resolution~\cite{Maletinsky2012, Grinolds2013}.
NV quantum sensing has been applied to various systems, from solid-state materials to biological macromolecules~\cite{DeLange2012, Hall2010, Shi2015, Grinolds2013, Lesage2013, Steinert2013, Abeywardana2016, Schirhagl2014}.
NV-detected electron paramagnetic resonance (EPR) from a single-electron spin has been demonstrated~\cite{Grinolds2014}.
Additionally, NV-detected nuclear magnetic resonance (NMR) with nanometer scale has been demonstrated~\cite{Waldherr2011, Mamin2013, Staudacher2013, Muller2014, Pham2016, Barry2020, doi:10.1073/pnas.2111607119}.

NMR is a powerful spectroscopic technique and is regularly used in various areas.
NV-detected NMR (NV-NMR) enables NMR spectroscopy with unprecedented signal sensitivity and a nanometer-/micrometer-scaled detection volume.
This unique capability will be useful for a wide variety of applications. 
Currently, most NV-NMR experiments are done at a low magnetic field of a few tens of milli teslas.
While NV-NMR at low magnetic fields is useful in many of those applications, high-field NV-NMR may be advantageous in some areas because of the higher resolution of chemical shifts, better signal sensitivity, and capability to observe a wider range of nuclei. 
For example, high-field NV-NMR can be useful in surface NMR for the study of chemically active sites at a molecular level, which often requires high sensitivity and the detection of nuclei with a low gyromagnetic ratio~\cite{TaChung17, doi:10.1021/jacs.9b13838, doi:10.1021/jacs.1c03162}.
NV-NMR at a high magnetic field is also advantageous in microfluidic NV-NMR using ensemble NV centers where the NMR signal intensity is proportional to the spin polarization of the detected nuclear spins~\cite{WalsworthNature2018}.
Moreover, high-field NV-NMR may be useful for the investigation of magnetic phases in solid states, which often requires the detection of nuclei with a low gyromagnetic ratio and benefit from high signal sensitivity and fine spatial resolution~\cite{Walstedt_spintemp, Walstedt_htc, Slichter, Yacoby2018}.
High Frequency (HF) NV-NMR experiments have been technically challenging.
In 2017, NV-NMR at 3 T was demonstrated~\cite{Aslam2017}.
We recently demonstrated NV-NMR at 8.3 T by employing electron–electron double resonance detected NMR (EDNMR) spectroscopy~\cite{Fortman2021}.
EDNMR spectroscopy is an EPR-based hyperfine spectroscopy in which the NMR signal is detected through changes in the population difference in the EPR transition with the application of a microwave pulse for a hyperfine-coupled transition, a so-called high turning angle pulse, to induce the population changes~\cite{Flores2008}.
Although EDNMR is a powerful technique due to its higher sensitivity to and resiliency against pulse-excitation-related artifacts, the NMR linewidth detected by EDNMR spectroscopy is limited by inhomogeneous broadening of the EPR transition, which is consequently much broader than the intrinsic NMR linewidth.
This is because the EPR (NMR) inhomogeneous broadening is proportional to the gyromagnetic ratio of electron (nuclear) spins.
In the previous EDNMR-based $^{13}$C NV-NMR experiment on diamond at 8.3 T, the observed NMR linewidth was $\sim 2.5$ MHz, which is much larger than the intrinsic $^{13}$C NMR linewidth.

In this work, we discuss the demonstration of HF NV-NMR at 4.2 T.
The technique is based on electron-nuclear double resonance (ENDOR) spectroscopy and is modified for the NV-NMR spectroscopy.
ENDOR is also an EPR-based hyperfine spectroscopy in which the NMR signal is detected through changes in the population difference in the EPR transition.
However, different from EDNMR used previously~\cite{Fortman2021}, ENDOR utilizes a radio-frequency (RF) pulse to excite the NMR transition which induces population changes in the EPR transition.
The duration of the NMR pulse can be extended as long as the $T_1$ relaxation time of the EPR transition.
Therefore, in principle, the linewidth of the present NV-NMR technique can be as narrow as an intrinsic NMR linewidth or Fourier-transform-limited linewidth of the NMR pulse if there are no other extrinsic contributions.
In the present experiment, we study HF $^{13}$C NV-NMR using two diamond crystals.
The first one is a type-Ib diamond, which is the diamond sample used in the previous study~\cite{Fortman2021}.
The other is a type-IIa diamond sample which has a narrower $^{13}$C NMR spectrum.
We study the $^{13}$C NV-NMR spectrum as a function of the duration of the RF pulse.
Then, by analyzing the NV-NMR spectra taken with various duration times, we extract components of the intrinsic linewidth and Fourier-transform-limited linewidth.
In the study of the type-Ib diamond (sample 1), we find that the linewidth of $^{13}$C NV-NMR becomes narrower as the duration of the RF pulse gets longer and that the linewidth is eventually limited by a full width at half maximum (FWHM) of 4 kHz, corresponding to a relative FWHM (FWHM divided by the NMR frequency) of 89 ppm.
This result demonstrates that the present method can improve the spectral resolution by three orders of magnitude compared with the previous investigation.
This method also measures the intrinsic $^{13}$C NMR of the type-Ib diamond sample.
In the study of the type-IIa diamond (sample 2), we show that the observed $^{13}$C NV-NMR is as narrow as 2 kHz, corresponding to a relative FWHM of 44 ppm.

\section{Method}
\begin{figure}
\label{HFsetup}
\centering
\includegraphics[width = 8 cm]{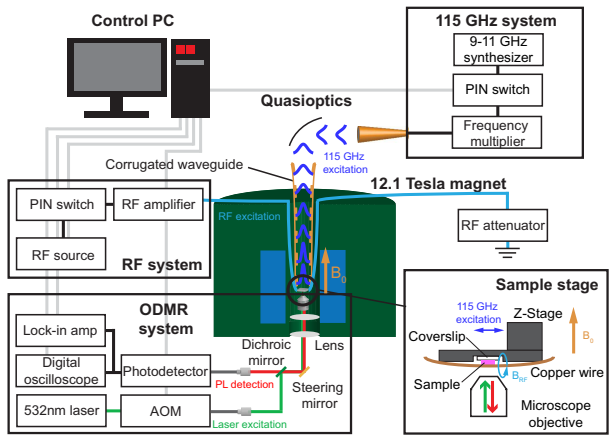}
\caption{Overview of the HF NV-NMR system.
HF microwaves are generated using the 115 GHz system.
The microwaves are guided through quasioptics before reaching the sample stage. The sample stage lies in the center of a 0–12.1 T superconducting magnet.
The ODMR system detects FL signals using a dichroic mirror and photodetector.
The RF system produces RF excitation which is sent through a copper wire on the surface of the sample.
The PC is responsible for controlling the instruments and experiment.
}
\label{fig:Setup}
\end{figure}
In the present investigation, we employ a home-built 115 GHz ODMR spectrometer.
Figure~\ref{fig:Setup} shows an overview of the HF ODMR system, which consists of a 115 GHz microwave source, quasioptics, a 12.1 T superconducting magnet, a sample stage, a RF system, an ODMR system, and a control PC.
The 115 GHz source (Virginia Diodes) outputs microwaves in the frequency range of 107 to 120 GHz (and also 215 to 240 GHz). The output power at 115 GHz is $\sim$400 mW.
In the present investigation, the magnetic field is set at $\sim$4.2 Tesla for the NV ODMR transition to be approximately 115 GHz.
These HF microwaves are then guided through quasioptics and a corrugated waveguide before reaching the sample stage.
The sample stage lies in the center of a superconducting magnet (Cryogenic Limited).
The magnetic field is continuously tunable between 0 and 12.1 T.
Within the sample stage, the sample is mounted onto a coverslip and a Z stage which can be used to adjust the height of the sample relative to the microscope objective.
Signals are detected using the ODMR system. which is based on a confocal microscope.
Photoluminescence (PL) signals of NV centers are detected by guiding the reflected light through a dichroic mirror, fluorescence filter, and detector.
An avalanche photodiode or photodiode is used for single or ensemble NV detection, respectively.
For the ensemble NV experiment, the photodiode is connected to a fast digital oscilloscope (Tektronix MSO64B) for pulse measurements~\cite{Li2021, Fortman2021}
or to a lock-in amplifier (Stanford Research, SR830) for continuous measurements.
HF and RF microwave excitations are controlled using a pulse generator (SpinCore Technologies PB-400) and the control PC.
Details of the HF ODMR spectrometer have been described previously~\cite{Cho2014,Cho2015, Stepanov2015, Fortman2020}.
Additionally, we added access for RF excitation in the HF ODMR system for NV-detected NMR experiments.
RF excitation is generated by the RF system, which consists of an RF source (Stanford Research, SR560), a pin switch (Mini Circuits ZASWA-2-50DR+), and an RF amplifier (TOMCO BT01000 GAMMA).
A copper wire with a diameter of $\sim$50 $\mu$m is placed tightly on the surface of the diamond for the coupling of RF excitation.

Two samples are studied in the present work.
Sample 1 is a $2.0 \times 2.0 \times 0.3$ mm$^{3}$, (111)-cut, high-pressure, high-temperature type-Ib diamond from Sumitomo Electric Industries.
The spin concentration of the single substitutional nitrogen defect (P1) centers is 10–100 ppm~\cite{KOLLARICS2022393, Fortman2021}.
The diamond sample was previously subjected to high-energy (4 MeV) electron beam irradiation and was exposed to a total fluence of $1.2 \times 10^{18}$ $e^{-}$/cm$^2$, followed by an annealing process at 1000 $^{\circ}$C.
This treatment produced a NV concentration greater than 1 ppm~\cite{Fortman2020, Fortman2019}.
Sample 2 is a $2.0 \times 2.0 \times 0.5$ mm$^{3}$, (100)-cut type-IIa diamond purchased from Element6 (DNV-B1).
The diamond sample was grown by chemical vapor deposition and is specified to have a NV concentration of $\sim$ 300 ppb and a P1 concentration of 800 ppb.
The investigations of both sample 1 and sample 2 were performed with measurements of ensemble NV centers.
Since the detection volume is $\sim$10 $\mu$m$^3$ in the present study, the number of detected NV centers is $> 2 \times 10^6$ and $\sim 5 \times 10^5$ for sample 1 and sample 2, respectively.

\section{Experiment and Results}
\begin{figure}
\centering
\includegraphics[width = 8 cm]{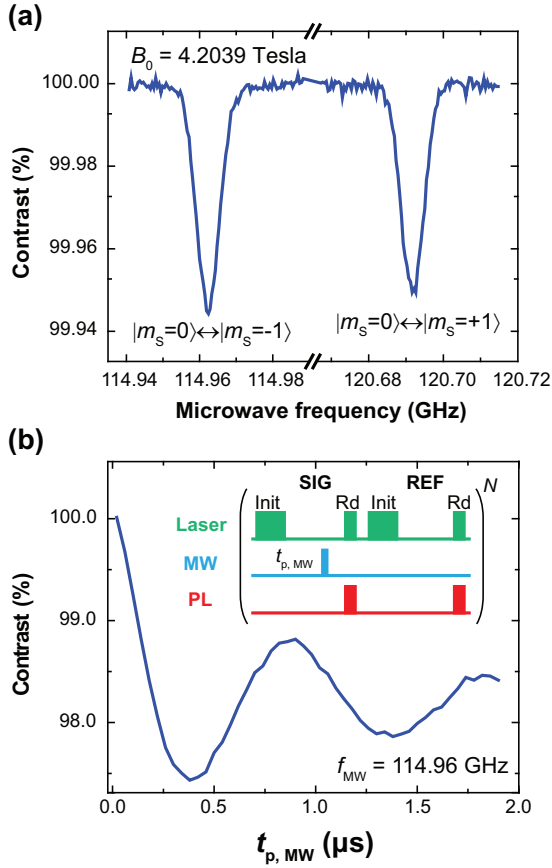}
\caption{Detection of ensemble NV centers in sample 1 at 4.2 T.
(a) NV ODMR spectrum. Two resonance dips were resolved clearly. The magnetic field and angle were obtained by analyzing the ODMR positions using the NV spin Hamiltonian.
(b) Rabi oscillations at a MW frequency of 114.96 GHz.
The inset shows a pulse sequence of the measurement.
The pulse sequence consists of "SIG" and "REF" sections.
The laser pulses of 700 and 1.28 $\mu$s were used for the initialization (Init) and the readout (Rd) of the NV states, respectively.
PL readout is achieved by selectively integrating the Rd part of the PL signal.
The sequence was repeated 9000 times for averaging ($N = 9000$.)
}
\label{characterization}
\end{figure}
The investigation of the HF NV-NMR linewidth started with NV ODMR measurements at 4.2 T.
The ODMR spectrum and Rabi oscillation data from sample 1 are shown in Fig.~\ref{characterization}.
As can be seen in Fig.~\ref{characterization}(a), the two peaks correspond to the transitions for $\ket{m_S = 0}$ $\leftrightarrow$ $\ket{m_S = +1}$ and $\ket{m_S = 0}$ $\leftrightarrow$ $\ket{m_S = -1}$ and are resolved at 114.96 and 120.69 GHz, respectively.
From the analysis of the ODMR frequencies, the magnetic field was determined to be B$_0$ = 4.2039 T with a tilt angle of $\theta$ = 3.5$^\circ$.
We then performed a Rabi oscillation measurement at the lower resonance frequency of 114.96 GHz.
The pulse sequence of the Rabi oscillation measurement is shown in the inset of Fig.~\ref{characterization}(b).
The sequence consists of two parts labeled "SIG" and "REF."
In SIG, the measurement starts from the initialization of the NV state into the $m_S=0$ state with the laser excitation.
Then, a microwave (MW) pulse with a duration of $t_{p,MW}$ is applied, and then the readout laser pulse is applied.
The MW frequency was fixed at 114.96 GHz.
The REF sequence contains the same laser initialization and readout pulses with no MW pulse applied and therefore corresponds to a PL intensity of $\ket{m_S = 0}$.
The sequence was repeated 9000 times for averaging.
Then, the contrast of the signal was obtained by calculating the ratio between SIG and REF for each $t_{p,MW}$ value, namely, SIG/REF.
This normalization using the REF signal allows us to reduce systematic noises significantly.
Figure~\ref{characterization}(b) shows the contrast with the experimental result.
We clearly observe Rabi oscillations.
From the values of $t_{p,MW}$ at the first dip, a $\pi$-pulse length of $t_{\pi, MW}=420$ ns was obtained.

\begin{figure}[ht]
\centering
\includegraphics[width = 8 cm]{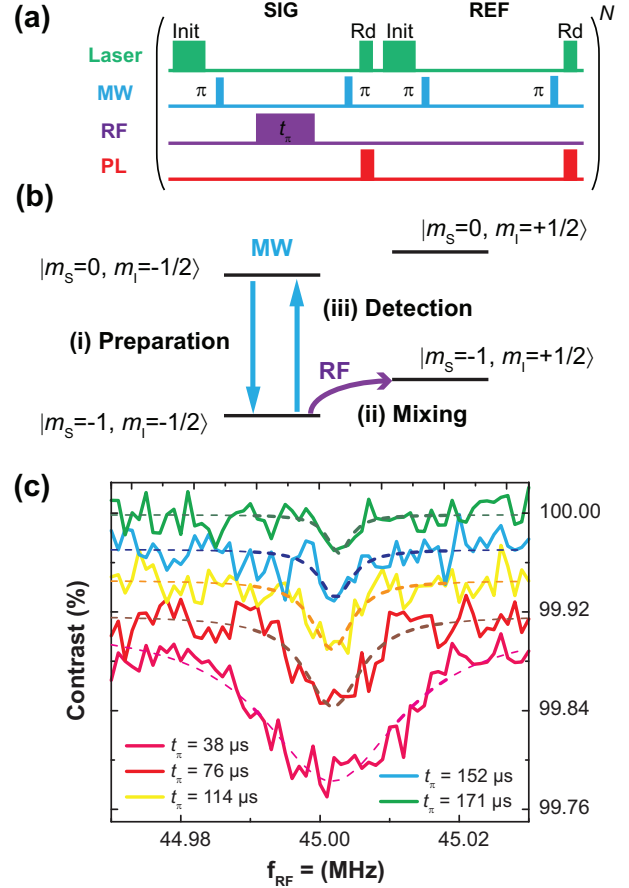}
\caption{$^{13}$C NV-NMR of sample 1 measured at 4.2 T.
(a) The modified Davies ENDOR sequence used for the NV-NMR detection.
MW is fixed on resonance with the NV transition, while the RF frequency is swept through the NMR transition.
The RF pulse length and power were adjusted to achieve close to a $\pi$-pulse rotation.
(b) Spin dynamics of the NV center during the NV-NMR sequence.
(c) $^{13}$C NV-NMR signals taken with various $\pi$-pulse lengths $t_{\pi}$ of the RF pulse. The power was adjusted to achieve close to a $\pi$ rotation for each measurement.
$N = 99000$.
}
\label{NVNMR_S1}
\end{figure}
After the characterization, we performed $^{13}$C NMR using NV centers.
We employed an ENDOR sequence modified from the Davis ENDOR~\cite{DAVIES19741} for NV-NMR.
The pulse sequence and the spin dynamics are shown in Figs.~\ref{NVNMR_S1}(a) and ~\ref{NVNMR_S1}(b).
This method essentially creates a hole in the NV ODMR spectrum.
As can be seen in Fig.~\ref{NVNMR_S1}(a), after laser initialization, the first MW $\pi$ pulse selectively drives the NV electron spin states from $\ket{m_S =0, m_I = -1/2}$ to $\ket{m_S =-1, m_I = -1/2}$ [(i) in Fig.~\ref{NVNMR_S1}(b)].
Then, the RF $\pi$ pulse is applied.
When the frequency of the RF pulse is on NMR resonance, the electron spin polarization is disturbed [(ii) in Fig.~\ref{NVNMR_S1}(b)].
Then, the second MW $\pi$ pulse drives the NV electron spin states back to $\ket{m_S =0, m_I = -1/2}$ for the readout of the NV states [(iii) in Fig.~\ref{NVNMR_S1}(b)].
On the other hand, the off-resonance RF pulse has no effect on the spin states.
In the measurement, the RF frequency is swept in the frequency range of the expected NMR transition. When the frequency of the RF pulse is on resonance, due to the population decrease of $\ket{m_S =0}$, a dip is seen in the PL intensity.
As we discuss later, the shape of the dip depends on the intrinsic NMR linewidth and the RF excitation bandwidth.
Therefore, in principle, it is possible to study the intrinsic linewidth of the NMR signal by reducing the RF excitation bandwidth.
The RF excitation bandwidth is determined by the length of the RF pulse.
The length of the RF pulse can be as long as $T_1$ of the NV centers (typically, a few milliseconds at room temperature)~\cite{Fortman2021}.
The experimental results for sample 1 can be seen in Fig.~\ref{NVNMR_S1}(c).
The signals were centered around 45.002 MHz, which agrees with the calculated value based on the ODMR data shown in Fig.~\ref{characterization}(a) and the $^{13}$C gyromagnetic ratio of $\gamma_C=10.705$ MHz/T.
As shown in Fig.~\ref{NVNMR_S1}(c), the NMR signal gets narrower when the RF pulse length gets longer.
The smallest FWHM was $4\pm2$ kHz, corresponding to a relative FWHM of 89 ppm.

\begin{figure}[ht]
\centering
\includegraphics[width = 8 cm]{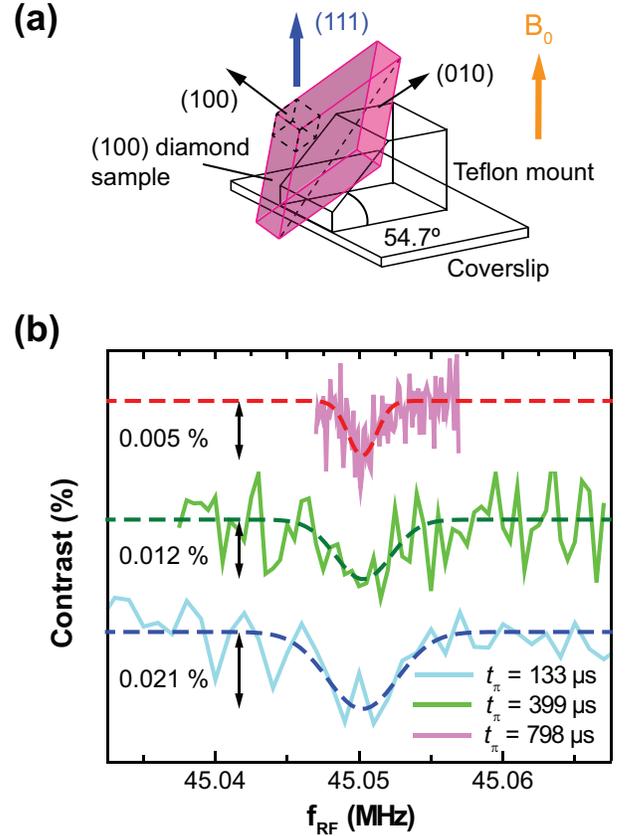}
\caption{$^{13}$C NV-NMR of sample 2.
(a) Schematic to show a method to mount a (100)-cut diamond.
The (100)-cut diamond sample is glued to a piece of Teflon angled at 54.7$^{\circ}$ to ensure that the {111} crystallographic orientation of NV centers is aligned with the magnetic field.
(b) NV-detected $^{13}$C NMR signals at 4.2 T with various $\pi$-pulse lengths $t_{\pi}$ of the RF pulse.
$N = 180000$ for the measurements with $t_{\pi} = 133$ and 399 $\mu$s. $N = 360000$ for the measurement with $t_{\pi} = 798$ $\mu$s.
From the ODMR spectrum analysis, we estimated the magnetic field strength to be 4.2107 T with a tilt angle of 8.3$^{\circ}$ in the experiment.
}
\label{NVNMR_dNV}
\end{figure}
Furthermore, we performed the HF NV-NMR experiment with sample 2.
Sample 2 has a smaller concentration of NV centers than sample 1.
Sample 2 is a (100)-cut diamond and therefore could not be mounted on a flat sample stage like sample 1.
To overcome this issue, we glued the sample to a piece of Teflon angled at 54.7$^{\circ}$. As can be seen in Fig.~\ref{NVNMR_dNV}(a), this ensured that the magnetic field was aligned with the (111) crystallographic orientation of the NV centers within the diamond.
As shown in Fig.~\ref{NVNMR_dNV}(b), using an RF pulse of $t_{\pi}=798$ $\mu$s, we observed a FWHM of 2$\pm$1 kHz, corresponding to a relative linewidth of 44 ppm.
The observed FWHM is half of the FWHM of sample 1.

\section{Discussion}
Here, we discuss the NMR linewidth observed in the present investigation.
As described in Fig.~\ref{NVNMR_S1}, the NMR signal originates from the excitation of the NMR transition caused by the RF pulse with a $\pi$-pulse length of $t_{\pi}$.
Therefore, the NMR linewidth depends on two contributions: the intrinsic NMR linewidth and the Fourier-transform-limited linewidth by $t_{\pi}$.
By considering spin dynamics of a two-level system, the Fourier-transform-limited linewidth is given by the probability of a NMR transition due to the RF pulse length.
This probability $P_{NMR}$ as a function of an RF pulse frequency $f_{RF}$ is given by~\cite{Stepanov2016}
\begin{multline}
P_{NMR}(\omega_{RF}; \omega_0, \omega_1, t_{\pi})  \\ =\frac{\omega_1^2}{(\omega_{RF}-\omega_0)^2 + \omega_1^2} \sin^2 \left(\frac{t_{\pi}}{2} \sqrt{(\omega_{RF}-\omega_0)^2 + \omega_1^2}\right),
\label{eq:trans}
\end{multline}
where $\omega_1 t_{\pi} = \pi$ and $t_{\pi}$ is the pulse length in units of seconds. $\omega_1 = \gamma_n B_1$ is the strength of an RF pulse ($B_1$), $\omega_{RF}$ is the frequency of the applied RF waves, and $\omega_0 = \gamma_n B_0$ is the Larmor frequency of NMR. $\omega_0$, $\omega_1$, and $\omega_{RF}$ are in units of radians per second.
When the RF pulse length is long enough, the experimental NMR linewidth will be equal to the intrinsic NMR linewidth.
In the present study, we consider a Lorentzian function to represent the intrinsic linewidth $L(\omega,\omega_c,\Delta\omega)$, namely,
\begin{equation}
    L(\omega,\omega_c,\Delta\omega) = \frac{1}{\pi} \frac{\Delta\omega/2}{(\omega - \omega_c)^2+ (\Delta\omega/2)^2},
\label{eq:intrinsic}
\end{equation}
where $\omega_c$ and $\Delta \omega$ are the center frequency and FWHM of the NMR signal, respectively.
Thus, by considering both contributions, the experimentally observed NMR signal $I_{NMR}$ can be modeled by the convolution of Eqs.(\ref{eq:trans}) and (\ref{eq:intrinsic}), namely,
\begin{multline}
I_{exp}(\omega_{RF}; \omega_0, \omega_1, \Delta\omega, t_{\pi})  \\ =\int P_{NMR}(\omega_{RF}; \omega', \omega_1, t_{\pi}) L(\omega', \omega_0, \Delta\omega) d\omega'.
\label{eq:total}
\end{multline}

Next, we discuss the intrinsic NMR linewidth by following the dipolar broadening theory shown by Van Vleck~\cite{PhysRev.74.1168} and Abragam~\cite{Abragam}. 
A NMR lineshape of nuclear spins in an inhomogeneous magnetic field has a certain width due to the spread of their Larmor frequencies.
In the case of a dilute spin system, the inhomogeneous magnetic field can originate from dipolar magnetic fields from surrounding nuclear and electron spins, which depend on the magnitude and orientation of their magnetic moments and the length and orientation of the vector describing their relative positions.
According to theory~\cite{PhysRev.74.1168, Abragam}, the NMR linewidth can be broaden and narrowed due to like-spin (e.g., $^{13}$C-$^{13}$C) and unlike-spin (e.g., $^{13}$C-P1) couplings. 
The linewidth can be calculated using the second and fourth moments of the line shape function ($M_2$ and $M_4$, respectively).
We start with the following equations for $M_2$ and $M_4$ influenced by the like-spin (denoted as "II") coupling (Eqs. (10) and (21) in Ref.~\cite{PhysRev.74.1168}). 
The second moment $M_2^{II}$ (in units of Hz$^2$) is given by
\begin{equation}
   h^2 M_2^{II}/(\frac{\mu_0}{4 \pi})^2 = \frac{I(I+1)}{3} \frac{1}{N} \sum_{j \neq k} B_{jk}^2,
   \label{Eq:M2_like}
\end{equation}
and the fourth moment $M_4^{II}$ (in units of Hz$^4$) is given by
\begin{equation}
\begin{split}
  & h^4 M_4^{II}/(\frac{\mu_0}{4 \pi})^4 
  = \{ \frac{I(I+1)}{3}\}^2 
   \frac{1}{N} \sum_{jkl \neq} \{3 B_{jk}^2 B_{jl}^2 
   + 2 A_{jk}^2 (B_{jl}-B_{kl})^2 \\
   & + 2 A_{jk} A_{kl} (B_{jl}-B_{jk})(B_{jl}-B_{kl})
 + 2 A_{jk} B_{jk} (B_{jl} - B_{kl})^2 \} \\
   & + \frac{2}{N} \sum_{k>j} [ B_{jk}^4 \frac{1}{5} \{ I^2(I+1)^2 -\frac{1}{3} I(I+1) \} \\
   & + 2 B_{jk}^3 A_{jk} \frac{1}{5} \{ \frac{2}{3} I^2(I+1)^2-\frac{1}{2}I(I+1) \} \\
   & + \frac{1}{2} B_jk^2 A_{jk}^2 \{\frac{4}{5} I^2(I+1)^2 - \frac{3}{5} I(I+1)\} ],
\end{split}
\label{Eq:M4_like}
\end{equation}
where $h$ is the Planck constant and $I$ is the nuclear spin number ($I=1/2$ for $^{13}$C). $N$ is the number of diamond lattice sites.
The notation $jkl \neq$ means that none of the indices are equal.
When the exchange coupling is negligible, $A_{jk} = -1/3 B_{jk}$. $B_{jk} = \gamma_I^2 h^2 b_{jk}$, where $b_{jk} = 3/2 (1-3 \cos{\theta_{jk}}^2)/r_{jk}^3$.
$\gamma_I$ is the gyromagnetic ratio of like spins (in hertz per tesla).
$r_{jk}$ is the distance between the $j$ and $k$ sites.
$\theta_{jk}$ is the angle between spin $j$ and spin $k$.
In addition, we define the concentration of the like spins $f_I$ as the probability to find a like spin at the $k$ site at a distance $r_{jk}$ from the occupied $j$ site. 
Using this definition, $\sum_{k \neq j} b^2_{jk} = N f_I \sum_{j} b^2_{jk}$~\cite{PhysRev.90.238}. 
Similarly, $\sum_{jkl \neq} b^2_{jk} b^2_{jl} = N f_I^2 \sum_{k \neq l} b^2_{jk} b^2_{jl}$, and $\sum_{k>j} b_{jk}^4 = N f_I/2 \sum_{k} b_{jk}^4$. 

Next, we consider the linewidths of both like-spin ($^{13}$C-$^{13}$C) and unlike-spin ($^{13}$C-P1) couplings.
When the system consists of unlike spins (P1 electron spins) in addition to like spins, the second and fourth moments are given by the sum of the like- and unlike-spin components, namely, 
$M_2 = M^{II}_2 + M^{IS}_2$ and 
$M_4 = M^{II}_4 + M^{IS}_4$,
where $M^{IS}_2$ and $M^{IS}_4$ are the second and fourth moments due to the dipolar couplings with unlike spins $S$, respectively.
According to Ref.~\cite{PhysRev.74.1168}, the second moment ($M_2^{IS}$) and fourth moment ($M_4^{IS}$) of an unlike-spin system are given by the following:
\begin{equation}
   h^2 M_2^{IS}/(\frac{\mu_0}{4 \pi})^2 = \frac{S(S+1)}{3} \frac{1}{N} \sum_{j \neq k'} C_{jk'}^2,
   \label{Eq:M2_unlike}
\end{equation}
and
\begin{equation}
\begin{split}
   & h^4 M_4^{IS}/(\frac{\mu_0}{4 \pi})^4 = \frac{I(I+1)}{3}\frac{S(S+1)}{3} 
   \frac{1}{N} \sum_{l'} \sum_{k>j} \{6 B_{jk}^2 (C_{jl'}^2 + C_{kl'}^2) \\
   &+ 2 A_{jk}^2 (C_{jl'}-C_{kl'})^2
   + 4 B_{jk} A_{jk} (C_{jl'} - C_{kl'})^2 \} \\
   & + (\frac{S(S+1)}{3})^2 \frac{1}{N} \sum_{j} \sum_{l'>k'} \{2 A_{k'l'}^2 (C_{jk'} - C_{jl'})^2+6 C_{jk'}^2 C_{jl'}^2\} \\
   & + \frac{1}{5} [S(S+1)-\frac{1}{3}] [S(S+1)] \frac{1}{N} \sum_{j,k'} C_{jk'}^4,
\label{Eq:M4_unlike}
\end{split}
\end{equation}
where the primed and unprimed sites are like spins ($^{13}$C) and unlike spins (P1 spins), respectively. 
$B_{jk} = \gamma_I^2 h^2 b_{jk}$, $B_{j'k'} = \gamma_S^2 h^2 b_{j'k'}$, and $C_{jk'} = 2/3 \gamma_I \gamma_S h^2 b_{jk'}$, where $b_{jk'} = 3/2 (1-3 \cos{\theta_jk'}^2)/r_{jk'}^3$.
$\gamma_S$ is the gyromagnetic ratio of unlike spins (in hertz per tesla).
Like in the previous section, we define the concentration of the unlike spins $f_S$ as the probability to find an occupied $k'$ site at a displacement $r_{jk'}$ from the occupied $j$ site.
$\sum_{j \neq k'} b^2_{jk'} = N f_s \sum_{k'} b^2_{jk'}$,
$\sum_{l'} \sum_{k>j} b^2_{jk}b^2_{jl'} = 1/2 \sum_{l'} \sum_{k \neq j} b^2_{jk}b^2_{jl'} = N f_I f_S/2 \sum_{j \neq l'} b^2_{jk}b^2_{jl'}$, 
$\sum_{j} \sum_{l'>k'} b^2_{k'l'}b^2_{jk'} 
= 1/2 \sum_{j} \sum_{k' \neq l'} b^2_{k'l'}b^2_{jk'} 
= N f_S^2/2 \sum_{k' \neq l'} b^2_{k'l'}b^2_{jk'}$,
and $\sum_{j, k'} b^4_{jk'} = N f_S \sum_{k'} b^4_{jk'}$. 
Moreover, for a dilute system ($f_I < 0.01$), $(M_2)^2/M_4 \ll 1/3$, and the truncated Lorentzian line shape is expected for the half-width at half maximum of the Lorentzian line ($\delta=\Delta\omega/2$); $M_2$ and $M_4$ are related as $M_2 = 2 \alpha \delta/\pi$ and $M_4 = 2 \alpha^3 \delta/(3\pi)$~\cite{Abragam}. 
$\alpha$ is a cut off parameter.
The FWHM of the NMR line shape function $\Delta\omega$ is given by 
\begin{equation}
   \Delta\omega = \frac{\pi}{\sqrt{3}} \sqrt{\frac{(M_2)^2}{M_4}} \sqrt{M_2}.
   \label{Eq:FWHM}
\end{equation}
In the present case, the FWHM is very sensitive to the P1 concentration.
Since $\gamma_S/\gamma_I$ is more than $10^3$ in the present case, 
as we can see from Eqs. (\ref{Eq:M2_like}) and (\ref{Eq:M2_unlike}) ($M_2^{II} \sim \gamma_I^4 f_I$ and $M_2^{IS} \sim \gamma_I^2 \gamma_S^2 f_S$), P1 spins can influence the FWHM of the $^{13}$C NMR spectrum when the P1 concentration $f_S$ is $10^6$ times more abundant than the $^{13}$C concentration $f_I$.

Finally, we evaluate $\Delta\omega$ for sample 1 and sample 2 numerically.
We start from the estimate of the FWHM of the $^{13}$C NMR linewidth with $I=1/2$, $\gamma_I=10.706$ MHz/T, and $f_I=0.011$.
The lattice constant of the diamond lattice is 3.567 {\AA}.
The volume used in the calculation is $7.1 \times 7.1 \times 7.1$ nm$^3$ ($21\times21\times21\times2=18522$ sites).
For the line broadening by like spins, using Eqs. (\ref{Eq:M2_like}) and (\ref{Eq:M4_like}), we obtain $M_2^{II} = 6.04 \times 10^4$ Hz$^2$ and $M_4^{II} = 4.71 \times 10^{10}$ Hz$^4$. 
In addition, $M_2^2/M_4 = 0.1 \ll 1/3$, which agrees with the Lorentzian line shape. 
Then, using Eq. (\ref{Eq:FWHM}), we obtain $\Delta\omega^{II} = 120$ Hz for the linewidth from the $^{13}$C-$^{13}$C coupling. 
This corresponds to a relative linewidth of 2.7 ppm at a $^{13}$C Larmor frequency of 45 MHz.
Next, we estimate $\Delta\omega$ of sample 1 and sample 2 using Eqs. (\ref{Eq:M2_unlike}), (\ref{Eq:M4_unlike}) and (\ref{Eq:FWHM}). 
For sample 1, we consider $f_S = 70$ ppm, $S=1/2$, and $\gamma_S = 28.024$ GHz/T.
It is worth noting that, in the calculation of the summation in Eqs. (\ref{Eq:M2_unlike}) and (\ref{Eq:M4_unlike}), the minimum distance was set to be 0.5 and 2 nm for the IS ($^{13}$C-P1) and SS (P1-P1) couplings, respectively, because couplings with closer distances will make those $^{13}$C and P1 spins energetically different from bulk $^{13}$C and P1 spins and their contributions to the linewidth are expected to be negligible.
As a result of the calculation, we obtain $M_2^{IS} = 3.05 \times 10^7$ Hz$^2$ $\gg M_2^{II}$ and $M_4^{IS} = 8.39 \times 10^{16}$ Hz$^4$ $\gg M_4^{II}$. 
Therefore, the $M_2$ and $M_4$ terms are dominated by the $M_2^{IS}$ and $M_4^{IS}$ terms. 
Also, $M_2^2/M_4 = 0.1 \ll 1/3$.
Using Eq. (\ref{Eq:FWHM}), we obtain $\Delta\omega = 1.1$ kHz (24 ppm), which shows significant line broadening from the linewidth by the $^{13}$C-$^{13}$C coupling ($\Delta \omega^{II}$).
The estimated FWHM value also agrees with the experimental value of a type-Ib diamond~\cite{doi:10.1021/acs.jpcc.2c06145}.
Next, we estimate the linewidths of sample 2.
With $f_S = 800$ ppb, we obtain $M_2^{IS} = 3.48 \times 10^5$ Hz$^2$ and $M_4^{IS} = 9.55 \times 10^{14}$ Hz$^4$, and $\Delta\omega = 15$ Hz (0.3 ppm) was obtained.
Like for sample 1, we found $M_2^{IS} \gg M_2^{II}$ and $M_4^{IS} \gg M_4^{II}$. However, different from sample 1, the estimated linewidth $\Delta\omega$ for sample 2 is smaller than $\Delta \omega^{II}$ (i.e., line narrowing). 

\begin{figure}[ht]
\centering
\includegraphics[width = 8 cm]{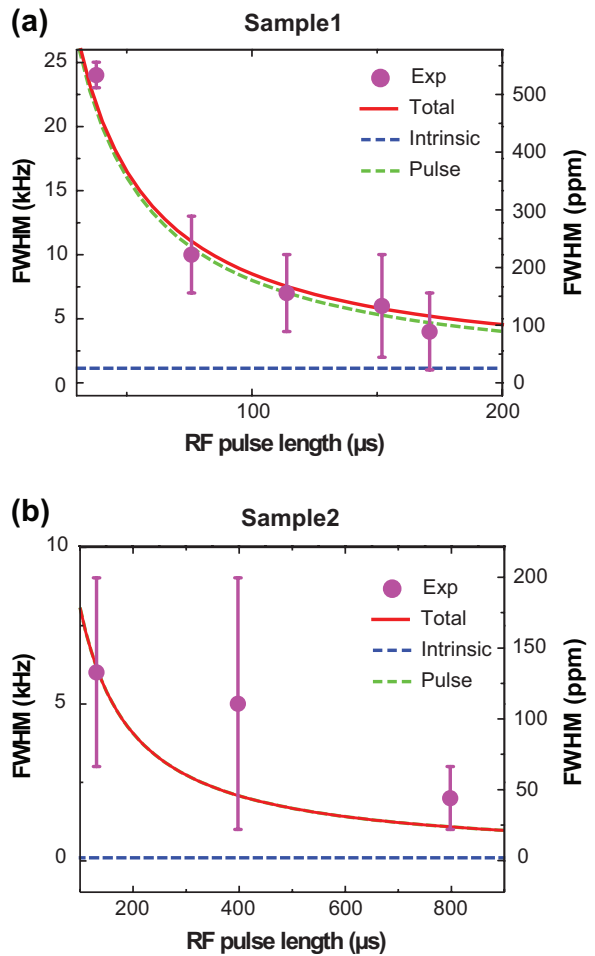}
\caption{NMR linewidth analysis for sample 1 and 2.
For each sample, the experimental FWHM values were extracted from the fit to the Lorentzian function.
(a) FWHM of the $^{13}$C NMR signal from sample 1 as a function of the RF pulse length.
The partial contributions from the NMR intrinsic linewidths ("Intrinsic") and the RF excitation ("Pulse") as well as the total linewidths ("Total") are shown by the dashed lines and solid line, respectively.
The error bars represent the 95$\%$ confidence interval.
(b) FWHM of the $^{13}$C NMR signal from sample 2 as a function of the RF pulse length.
}
\label{linewidth_analysis}
\end{figure}
Finally, we discuss the contributions to the observed NMR linewidth.
For both samples, the FWHM values were determined by fitting the experimental NV-NMR data with the Lorentzian function.
Figure \ref{linewidth_analysis} shows the experimentally determined FWHM and an analysis of the contributions from the intrinsic NMR linewidth and the excitation bandwidth from the RF pulse.
The RF excitation bandwidth is calculated using Eq. (\ref{eq:trans}) with the given $t_{\pi}$ ["Pulse" in Figs.~\ref{linewidth_analysis}(a) and \ref{linewidth_analysis}(b)].
The intrinsic linewidth is $\Delta \omega$ in Eq. (\ref{eq:intrinsic}).
For the analysis of sample 1, we set $\Delta \omega = 1.1$ kHz by considering the contribution from the P1 centers with a concentration of 70 ppm ["Intrinsic" in Fig.~\ref{linewidth_analysis}(a)].
On the other hand, for the analysis of sample 2, we set $\Delta \omega = 15$ Hz ["Intrinsic" in Fig.~\ref{linewidth_analysis}(b)].
Then, the convoluted FWHM values are calculated using Eq. (\ref{eq:total}) ["Total" in Figs.~\ref{linewidth_analysis}(a) and \ref{linewidth_analysis}(b)].
As can be seen from Fig.~\ref{linewidth_analysis}(a), the observed FWHM values from sample 1 agree well with the simulation, showing that the FWHM value is limited by the excitation bandwidth for the measurement with short RF pulses and by the intrinsic NMR linewidth for long pulse measurements.
As shown in Fig.~\ref{linewidth_analysis}(b), the observed FWHM values of sample 2 and the analysis agree fairly.
The analysis indicates that a longer RF pulse may be able to improve the spectral resolution more to observe the intrinsic NMR linewidth; however, such a measurement may be challenging because the length of the RF pulse is limited by the $T_1$ relaxation time of the NV centers.

\section{Summary}
In summary, we successfully demonstrated NV-detected NMR of $^{13}$C nuclear spins at 4.2 T using a modified Davies ENDOR sequence.
We implemented RF access in the HF NV-NMR system for the present study.
The demonstration showed that this ENDOR-based NV NMR is capable of obtaining the intrinsic $^{13}$C NMR from sample 1.
Using a type-IIa diamond crystal (sample 2), we were able to experimentally obtain a narrower linewidth closer to the intrinsic linewidth of the $^{13}$C NMR signal.
In this work, we have developed a technique that opens the door to further studies of NV-detected NMR at even higher magnetic fields.
The demonstrated NV-NMR technique, capable of detecting a linewidth of 44 ppm, can be applied to other investigations, including NMR studies of magnetism in solids.
Furthermore, the presented NV-NMR technique can be advanced further with dynamic nuclear polarization~\cite{doi:10.1021/acs.jpcc.2c06145} and advanced pulsed NV-NMR methods to improve the signal sensitivity and the spectral resolution. 
Future studies of NV-NMR at high magnetic fields may include surface NMR and microfluidic NMR at high magnetic field.

\section{Acknowledgements}
This work was supported by the National Science Foundation (Grants No. ECCS-2204667 and No. CHE-2004252, with partial co funding from the Quantum Information Science program in the Division of Physics), the USC Anton B. Burg Foundation, and the Searle Scholars Program (S.T.).

%

\end{document}